\pdfoutput=1
\documentclass[10pt,twocolumn,letterpaper,]{article}
\usepackage[T1]{fontenc}
\usepackage[utf8]{inputenc}
\usepackage{authblk}
\usepackage[pagenumbers]{cvpr} 
\usepackage{array} 
\usepackage{pifont} 
\usepackage[dvipsnames]{xcolor}

\definecolor{cvprblue}{rgb}{0.21,0.49,0.74}
\usepackage[pagebackref,breaklinks,colorlinks,citecolor=cvprblue]{hyperref}
\usepackage{multirow}
\usepackage[utf8]{inputenc}
\usepackage[margin=1in]{geometry} 
\usepackage{xcolor}
\usepackage{tcolorbox}
\usepackage{enumitem}

\definecolor{bg-gray}{RGB}{242, 242, 242}
\usepackage[table]{xcolor}
\usepackage{booktabs}   
\usepackage{multirow}   
\usepackage{amsmath}    
\usepackage[normalem]{ulem} 
\usepackage{booktabs}
\usepackage{multirow}
\usepackage{amsmath}
\usepackage{graphicx}  
\usepackage{float}     
\usepackage{graphicx}
\usepackage{lipsum}
\usepackage{caption}
\usepackage{float}
\usepackage{amsmath}
\usepackage{algorithm}
\usepackage{algorithmic}
\usepackage[table]{xcolor}  
\usepackage{amssymb}      
\usepackage{booktabs}
\usepackage{multirow}
\usepackage{amsmath}
\usepackage[table]{xcolor}  
\usepackage{amssymb}      
\usepackage{booktabs}
\usepackage{multirow}
\usepackage{amsmath}
\usepackage{siunitx}      
\usepackage{hyperref}
\hypersetup{linkcolor=cvprblue,urlcolor=cvprblue,filecolor=cvprblue,citecolor=cvprblue}
\def\paperID{} 
\def\confName{CVPR}
\def\confYear{2026}

\author{
\vspace{-2em}
    Yusheng Dai$^{1,*}$, Kangdi Wang$^{2,*}$, Baolong Gao$^{3}$, Yuxuan Jiang$^{3}$, \\
    \vspace{-0.8em} Weiqiang Wang$^{1}$, Qiuhong Ke$^{1}$, Jianfei Cai$^{1}$ \\
    $^1$ Monash University \quad $^2$ University of Chinese Academy of Sciences \quad $^3$ Tsinghua University
}
\vspace{-2em}
\title{CineDub: Scaling End-to-End Video Dubbing to Multi-Speaker Dialogues with Coherent Sound Effects}

\usepackage{makecell}
\usepackage{subcaption}

\providecommand{\Description}[1]{}
\providecommand{\authornote}[1]{}
\providecommand{\authornotemark}[1][]{}
\providecommand{\orcid}[1]{}
\providecommand{\affiliation}[2][]{}

\providecommand{\setcopyright}[1]{}
\providecommand{\copyrightyear}[1]{}
\providecommand{\acmYear}[1]{}
\providecommand{\acmDOI}[1]{}
\providecommand{\acmConference}[4][]{}
\providecommand{\acmBooktitle}[1]{}
\providecommand{\acmISBN}[1]{}
\providecommand{\acmSubmissionID}[1]{}
\providecommand{\ccsdesc}[2][]{}
\providecommand{\keywords}[1]{}
\providecommand{\received}[2][]{}
\providecommand{\settopmatter}[1]{}
\providecommand{\citestyle}[1]{}
\providecommand\footnotetextcopyrightpermission[1]{}
\usepackage{verbatim}

\begin{document}

\setlength{\abovecaptionskip}{5pt}
\setlength{\textfloatsep}{5pt}
\setlength{\intextsep}{-5pt}

\maketitle
\vspace{-1.8em}

\begin{abstract}
\vspace{-10pt}
\begingroup
\renewcommand\thefootnote{}
\footnotetext{$^{*}$Equal contribution.}
\endgroup
Automatic video dubbing in the wild remains fundamentally limited by two competing constraints: hierarchical methods depend on brittle, multi-stage preprocessing pipelines that severely restrict data scalability and practical deployment, while holistic approaches operating on uncropped video suffer from weak temporal alignment and speaker-utterance ambiguity in multi-speaker settings. To overcome these limitations, we propose \textbf{CineDub}, a unified diffusion-based model that achieves precise multi-speaker dialogue dubbing directly from uncropped videos, without face cropping or speaker diarization. Central to our approach is the Implicitly-Coupled Holistic Conditioning \textbf{(ICHC)} paradigm, where holistic visual representations and a semantic-bundled transcription format are encoded independently, yet implicitly coupled through cross-modal training to resolve speaker ambiguity and enable precise multi-speaker multi-turn dialogue dubbing. Building on the unified temporal cues captured by holistic visual features, we further extend CineDub to joint speech and audio generation. We introduce an Ambient-to-Linguistic Curriculum Learning \textbf{(ALC)} to mitigate sub-task degradation, and a decoupled textual branch control mechanism to resolve cross-prompt interference during simultaneous generation. We also release two in-the-wild benchmarks, \textbf{CineDub-Multi} for multi-speaker dialogue dubbing and \textbf{CineDub-SA} for video-to-speech-and-audio (V2SA) generation, to enable evaluation under realistic conditions. Experiments show that CineDub achieves state-of-the-art results on established single-speaker dubbing and video-to-audio benchmarks while excelling in multi-speaker dialogue dubbing and acoustically coherent joint generation.
\end{abstract}
\vspace{-10pt}

\section{Introduction}

\begin{figure}[t]
  \centering
  \includegraphics[width=\columnwidth]{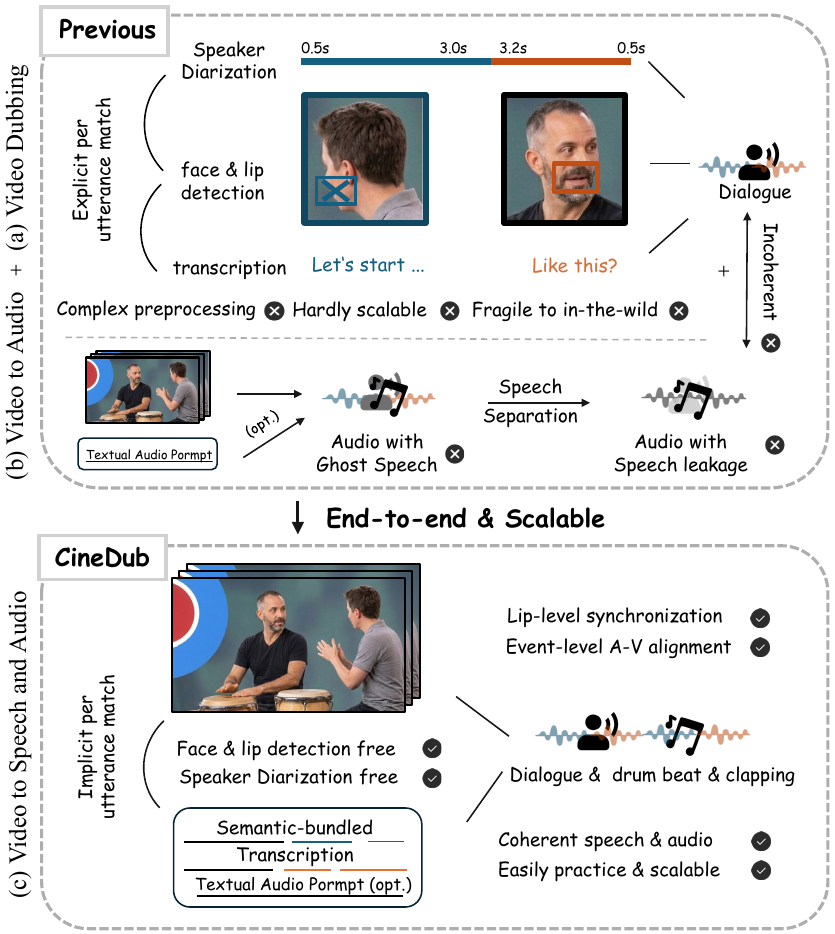}
  \caption{(a)~Existing dialogue dubbing methods rely on complex preprocessing to provide active
  cropped speaker regions with corresponding timestamps and transcripts, limiting scalability and practicality. (b)~Generating accompanying sound effects requires a separate V2A model, which inevitably introduces ghost speech and acoustic incoherence with the dubbed speech. (c)~CineDub jointly generates multi-speaker dialogue with coherent audio through a unified end-to-end model, requiring only an uncropped video and an MLLM-generated semantic-bundled transcription.}
  \Description{Comparison of three approaches: (a) per-utterance preprocessing pipeline for multi-speaker dubbing, (b) separate V2A model with ghost speech artifacts, and (c) CineDub's end-to-end joint speech-and-audio generation from uncropped video.}
  \label{fig:teaser}
\end{figure}

\label{sec:intro}
Automatic video dubbing synthesizes target speech from a given transcript, a reference speaker video, and an optional reference speech. Unlike conventional text-to-speech, this task requires the synthesized speech to align with visual cues in both lip movements and prosodic expressiveness, making it more challenging yet highly valuable for multimedia production. \footnote{\href{https://cinedub2026.github.io/}{https://cinedub2026.github.io}}

Most existing video dubbing methods follow a hierarchical paradigm~\cite{cong2023hpmdubbing,cong2024styledubber,choi2025aligndit,cong2026cosyncdit,ye2026hierarchical} that extracts visual features at multiple granularities (e.g., lip or face crops) to improve lip synchronization and speech expressiveness. While effective on single-speaker benchmarks, these methods depend on multi-stage preprocessing pipelines and heterogeneous pre-trained encoders, making them unwieldy in practice and brittle in challenging scenarios (Figure~\ref{fig:teaser}a). Although recent open-source toolkits~\cite{liu2026funcineforge} offer a more engineering-friendly workflow and extend to multi-speaker scenarios, the additional preprocessing stages they introduce (e.g., active speaker detection and speaker diarization) further increase complexity and remain fragile under realistic conditions. More critically, at training time, the costly preprocessing severely complicates training data curation, making this paradigm hardly scalable to in-the-wild scenarios.

Recent methods~\cite{zheng2025deepdubber,zhang2025deepaudiov1} take a different approach, extracting unified visual control signals directly from uncropped videos to obtain simpler and more generalizable architectures. However, these holistic approaches share two critical weaknesses. First, without explicit local spatial supervision, they lack the fine-grained temporal cues needed for precise lip-speech synchronization. Second, the holistic representation introduces \textit{speaker-utterance ambiguity} in multi-speaker scenarios: visual features of all speakers are entangled within a single representation, leaving the model unable to assign utterances to their corresponding speakers or determine correct turn-taking. This problem is compounded by frequent shot changes and off-screen cuts, which further blur the visual-speech correspondence. Thus, achieving both precise temporal alignment and unambiguous speaker assignment within a holistic framework remains an open challenge.

To address these challenges, we present CineDub, a unified diffusion-based model for video dubbing that operates on holistic video and achieves robust performance in complex multi-speaker dialogue scenarios (Figure~\ref{fig:teaser}c). The core contribution of CineDub lies in the \textit{Implicitly-Coupled Holistic Conditioning} (ICHC) paradigm, where holistic visual and textual conditions are encoded individually to ensure simplicity and scalability, yet implicitly coupled through cross-modal training to resolve speaker-utterance ambiguity. On the visual side, we find that SynchFormer~\cite{iashin2024synchformer} features, traditionally used for event-level audio-visual association in video-to-audio generation, can also capture fine-grained lip synchronization cues for video dubbing (Figure~\ref{fig:synch_attn}). More surprisingly, in multi-turn dialogues with multiple speakers, SynchFormer exhibits an emergent attention-switching behavior without active speaker detection. It dynamically shifts focus to the active speaker as turns change, even when multiple speakers are visible in the same frame. On the textual side, to resolve the speaker-utterance ambiguity introduced by the holistic visual condition, we design a \textit{semantic-bundled transcription format} that couples each segment-level speaker description with its corresponding transcript segment in chronological order. This format acts as an implicit prompt that guides the model to associate each transcription segment with the temporal visual trace of the correct speaker, resolving the ambiguity. Since this structured transcription can be directly generated by MLLMs, it scales easily to large in-the-wild datasets without complex speech frontend pipelines, while also offering users segment-level control over each speaker's emotion and timbre.

Furthermore, the holistic paradigm naturally extends to end-to-end joint speech and audio generation,\footnote{In this paper, \textit{audio} and \textit{sound} refer to any non-speech sound such as soundscapes and event sounds. \textit{Speech generation} refers to transcript-conditioned synthesis. \textit{Video-to-audio} (V2A) subsumes both video-to-audio and video-text-to-audio generation.} bypassing the cascaded pipelines that generate speech and audio separately and inevitably introduce ghost speech artifacts and acoustic incoherence (Figure~\ref{fig:teaser}b). We observe that SynchFormer already provides a unified temporal visual condition that captures both event-level audio-visual correspondences and fine-grained lip synchronization cues (Section~\ref{sec:visual_condition}), enabling a single model to handle both tasks jointly.
However, achieving high-fidelity joint generation under this holistic paradigm remains unexplored and faces two challenges: sub-task performance degradation and cross-prompt interference. To address sub-task degradation, we introduce an \textit{Ambient-to-Linguistic Curriculum Learning} (ALC). Inspired by human linguistic evolution, the model first learns from general audio and gradually specializes toward symbolic human speech. This ordering resolves the optimization conflict between speech and audio under shared visual conditioning, enabling effective joint training without degrading either subtask. Second, during simultaneous generation, injecting both the audio and transcription prompts through a shared attention module often leads to severe attention dilution. The model tends to disproportionately suppress the audio prompt, resulting in poor instruction adherence for sound effects. We mitigate this interference via a \textit{decoupled textual branch control mechanism} that routes each prompt through independent cross-attention branches, with learnable meta-tokens replacing inactive branches during single-task inference to eliminate cross-task leakage.

Finally, to address the limitations of existing benchmarks in real-world scenarios, we release two in-the-wild benchmarks (Section~\ref{sec:benchmark}): \textit{CineDub-Multi} for multi-speaker dialogue dubbing with co-occurring speakers, moving beyond the single-speaker assumption of prior benchmarks; and \textit{CineDub-SA} for V2SA evaluation, comprising 10-second clips from VGGSound~\cite{chen2020vggsound} with curated on-screen talking faces that ensure verified audio-visual correspondence and stable evaluation via existing in-domain embedding metrics.

In summary, our contributions are as follows:
\begin{itemize}
\item We propose CineDub, an end-to-end diffusion video dubbing framework that operates directly on uncropped videos. The novel Implicitly-Coupled Holistic Conditioning (ICHC) paradigm implicitly couples holistic visual features with a semantic-bundled transcription format, ensuring precise multi-speaker multi-turn dialogue dubbing. By bypassing brittle preprocessing pipelines,
CineDub greatly simplifies data curation and scales to diverse in-the-wild scenarios.

\item We extend the ICHC paradigm to unified speech and audio joint generation, and identify two design principles: an Ambient-to-Linguistic Curriculum Learning (ALC) that first builds a broad audio prior before specializing to speech, and a decoupled textual branch control mechanism that routes heterogeneous conditions through independent branches to prevent cross-prompt interference.

\item We construct two in-the-wild benchmarks, CineDub-Multi and CineDub-SA, to enable evaluation under realistic multi-speaker dialogue and joint speech and audio settings. Experiments show that CineDub achieves state-of-the-art performance across standard benchmarks while scaling to cinematic scenarios with multiple speakers and sound effects.
\end{itemize}

\section{Related Work}

\subsection{Automatic Video Dubbing}

Most video dubbing methods follow a hierarchical paradigm. HPMDubbing~\cite{cong2023hpmdubbing} uses lip crops for temporal alignment, face crops for emotional prosody~\cite{wang2024emotion}. More recently, AlignDiT~\cite{choi2025aligndit} unified text, video, and audio within a DiT backbone, yet still requires pre-cropped mouth regions. To handle multi-speaker scenarios, FunCineForge~\cite{liu2026funcineforge} further introduces active speaker detection and speaker diarization, though the added complexity remains fragile under realistic conditions. Alternatively, recent methods derive visual control signals directly from uncropped video. DeepDubber~\cite{zheng2025deepdubber} and InstructDubber~\cite{zhang2026instructdubber} leverage MLLMs to infer semantic attributes that enhance prosody control. DeepAudio~\cite{zhang2025deepaudiov1} and DualDub~\cite{tian2025dualdub} forgo explicit reasoning and generate speech directly from holistic visual features via implicit alignment. However, these methods either lack fine-grained temporal cues for lip synchronization, or entangle all speakers into a single representation, failing to resolve speaker-utterance ambiguity.
\vspace{5pt}
\subsection{{\fontsize{10.5}{13}\selectfont\bfseries Video to Speech and Audio Joint Generation}}
Most systems generate speech and audio separately and combine them via linear superposition~\cite{zhang2025deepaudiov1,zhang2025lvas}. DeepAudio~\cite{zhang2025deepaudiov1} first produces sound effects with a V2A model and uses the resulting energy as a prior to guide speech synthesis. However, cascaded pipelines inherently suffer from two problems: \textit{acoustic incoherence} between independently generated speech and audio, and \textit{ghost speech artifacts} where the V2A model produces spurious speech-like sounds that corrupt the final mix. Although LVAS-Agent~\cite{zhang2025lvas,song2025hume} further introduces a Synthesizer Agent, the fundamental limitations still persist. Joint generation within a single model offers a promising alternative. Among discrete-token methods, DualDub~\cite{tian2025dualdub} and BVS~\cite{niu2025bvs} jointly decode speech and audio tokens. However, single-pass discrete decoding precludes iterative visual-conditioned refinement, yielding suboptimal audio-visual alignment. On the diffusion side, AudioGen-Omni~\cite{wang2025audiogenomni} shows strong potential with a multimodal DiT but is bottlenecked by scarce joint training data. VSSFlow~\cite{cheng2025vssflow} mitigates this via feature-space synthesis but relies on cropped lip video with separate visual conditions. For subsequent developments after our submission to the target venue, we point readers to recent efforts~\cite{tao2026foleyomni,guan2026holidubber,pian2026omnisonic}. All above joint generation systems, however, remain limited to single-speaker scenarios, leaving multi-speaker dialogue dubbing unaddressed.

\subsection{Benchmarks on Video Dubbing and V2SA}
\label{sec:related_benchmarks}
Most existing public video dubbing benchmarks target single-speaker scenarios~\cite{wang2025audioatlas}; while FunCineForge~\cite{liu2026funcineforge} recently released the first multi-speaker Chinese television dubbing benchmark, a comparable English benchmark remains absent. For V2SA evaluation, two public benchmarks exist. DualBench~\cite{tian2025dualdub}, drawn from the V2C-Animation dataset~\cite{chen2022v2c} of Disney animated films, is not publicly available due to copyright restrictions. Moreover, its clips are only 2--3 seconds long, too short for embedding-based audio metrics trained on 10-second audio. Its sound-effect categories are also biased toward cartoon-style foley. AC-filtered~\cite{kim2019audiocaps} was originally developed for text-to-speech-and-audio generation. When repurposed for V2SA evaluation, a large portion of its audio consists of off-screen narration with weak audio-visual correspondence, making it unsuitable for assessing audio-visual synchronization. Thus, a reliable V2SA benchmark is still lacking.

\section{Method}

\subsection{Preliminaries}

\subsubsection{Diffusion-based Audio Generation}
CineDub is built on a latent diffusion framework with a fully-convolutional variational autoencoder (VAE) that operates directly on 16k raw waveforms~\cite{evans2024stableaudioopen,dai2025latentswap}. Following the Descript Audio Codec~\cite{kumar2023dac} architecture but removing the quantization bottleneck, this VAE compresses audio into a continuous latent representation $x_0 \in \mathbb{R}^{T_a \times D_a}$, where $T_a$ is the temporal length and $D_a$ is the latent dimension. The diffusion process progressively corrupts $x_0$ by adding Gaussian noise according to a variance schedule $\{\bar{\alpha}_t\}_{t=1}^{T}$:
\begin{equation}
q(x_t | x_0) = \mathcal{N}(x_t; \sqrt{\bar{\alpha}_t}\, x_0, (1 - \bar{\alpha}_t)\, \mathbf{I}),
\label{eq:forward}
\end{equation}
where $t$ is the diffusion timestep. A denoising network $\epsilon_\theta$ is trained to predict the noise $\epsilon$ given the noisy latent $x_t$ and conditioning signals $c$:
\begin{equation}
\mathcal{L}_{\text{diff}} = \mathbb{E}_{t, \epsilon, x_0} \left\| \epsilon_\theta(x_t, c, t) - \epsilon \right\|^2.
\label{eq:diffloss}
\end{equation}

We adopt a Diffusion Transformer (DiT)~\cite{peebles2023dit} as the denoising backbone. Each DiT block processes the latent sequence $x_t$ and incorporates conditions through two mechanisms: cross-attention for semantically-rich conditions, and channel-wise concatenation followed by self-attention for temporally-dense conditions~\cite{jiang2025freeaudio}. CineDub extends this framework to handle heterogeneous multi-modal inputs for joint speech and audio generation. Following prior work~\cite{choi2025aligndit}, we adopt a speech infilling formulation for voice cloning: during training, a clean reference speech latent is concatenated as a prefix to the noisy target latent, and the model learns to reconstruct only the target portion, enabling zero-shot timbre transfer via in-context learning.

\subsubsection{Audio-Visual Temporal Synchronization}
\label{sec:synchformer_prelim}
SynchFormer~\cite{iashin2024synchformer} is a self-supervised audio-visual synchronization model widely used in video-to-audio generation~\cite{cheng2025mmaudio,bain2025vintage} for extracting temporally-aligned visual features. It divides an input video into $S$ equally-spaced segments and extracts segment-level features with a visual encoder (Motionformer~\cite{patrick2021motionformer}) and an audio encoder (AST~\cite{gong2021ast}).  Training of SynchFormer proceeds in two stages. First, segment-level contrastive pre-training (Segment AVCLIP) aligns audio and visual features within each segment:
\begin{equation}
\mathcal{L}_{\text{AVCLIP}} = -\frac{1}{BS}\sum_{i=1}^{BS} \log \frac{\exp(\tilde{a}_i \cdot \tilde{v}_i / \tau)}{\sum_{j=1}^{BS} \exp(\tilde{a}_i \cdot \tilde{v}_j / \tau)},
\label{eq:avclip}
\end{equation}
where $\tilde{a}_i$ and $\tilde{v}_i$ are audio and visual features from segment $i$, $B$ is the batch size, and $\tau$ is a learnable temperature. Second, features from all segments are concatenated and passed to a lightweight transformer that classifies the temporal offset, trained with cross-entropy loss while keeping the encoders frozen. The entire pipeline is trained on AudioSet~\cite{gemmeke2017audio}, a large-scale in-the-wild dataset of videos with both sound and speech. In CineDub, we extract the segment-level visual features from the frozen first-stage encoder and concatenate them along the temporal dimension to obtain $\mathbf{c}_v \in \mathbb{R}^{T_v \times D_v}$, which serves as the holistic visual condition.

\subsection{Implicitly-Coupled Holistic Conditioning}
\label{sec:ichc}

\begin{figure}[t]
  \centering
  \includegraphics[width=\linewidth]{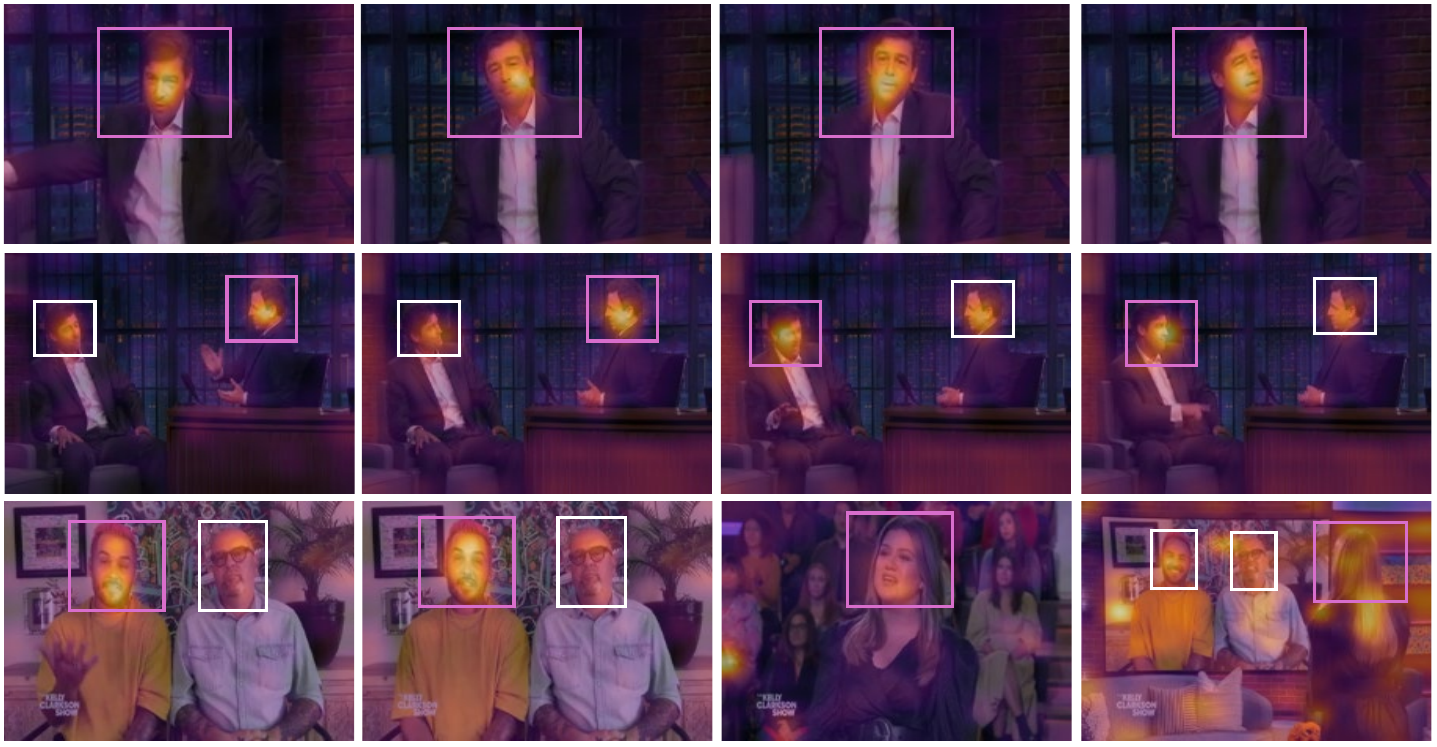}
  \caption{Visualization of SynchFormer self-attention maps (magenta: active speaker; white: inactive speaker). (a)~Single-speaker setting: attention concentrates on the lip region. (b)~Two-speaker setting: attention dynamically shifts to the active speaker across turns. (c)~Failure cases: attention drifts to a non-speaking face during overlapping speech or becomes ambiguous at shot boundaries.}
  \Description{Visualization of SynchFormer self-attention maps showing lip-focused attention in single-speaker settings, attention switching between speakers in two-speaker dialogues, and attention drift failure cases in multi-speaker scenarios.}
  \label{fig:synch_attn}
\end{figure}

\begin{figure*}[t]
  \centering
  \includegraphics[width=1.0\linewidth]{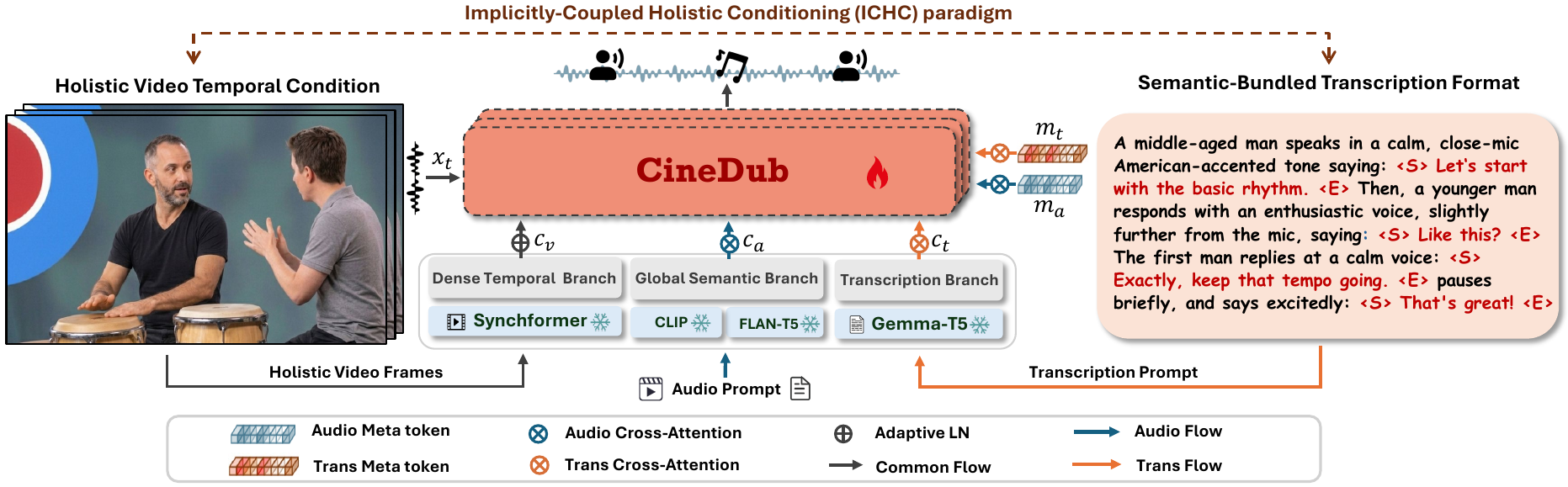}
  \caption{Overview of CineDub under the ICHC paradigm. The holistic visual condition $\mathbf{c}_v$ (left), extracted by SynchFormer (Sec.~\ref{sec:visual_condition}), encodes both event-level audio-visual correspondences and fine-grained lip-sync alignment. To address the speaker assignment ambiguity in $\mathbf{c}_v$, the semantic-bundled transcription $\mathbf{c}_t$ (right) provides per-segment speaker-utterance grounding cues, implicitly coupling with $\mathbf{c}_v$ via multi-conditional training (Sec.~\ref{sec:text_condition}). CineDub adopts a unified DiT backbone that supports both joint video-to-speech-and-audio generation and each subtask. $\mathbf{c}_t$ and $\mathbf{c}_a$ are routed through decoupled textual branches to prevent cross-prompt interference, with learnable meta-tokens replacing inactive branches during single-task inference (Sec.~\ref{sec:decoupled_attn}).}
  \Description{Framework overview of CineDub showing three conditioning branches (SynchFormer, CLIP+FLAN-T5, Gemma-T5) feeding into a diffusion DiT backbone with decoupled cross-attention, meta-tokens, and task-specific flow paths for joint speech-and-audio generation.}
  \label{fig:framework}
\vspace{-5pt}
\end{figure*}

\subsubsection{Holistic Visual Condition}
\label{sec:visual_condition}

Given that SynchFormer provides dense, temporally-aligned visual features for V2A generation, we first explore whether these features can also supply fine-grained lip synchronization cues for video dubbing. We visualize the self-attention of SynchFormer's MotionFormer visual encoder using Attention Rollout~\cite{abnar2020quantifying} in Figure~\ref{fig:synch_attn}. Across diverse scenarios we observe two notable properties: (1)~in single-speaker settings (Figure~\ref{fig:synch_attn}a), SynchFormer concentrates attention around the face of the speaker, particularly the lip region—despite being trained only with a general audio-visual synchronization objective, it learns to localize lips as the most temporally informative cue for speech, providing the fine-grained alignment that dubbing requires; (2)~when multiple speakers appear in the same frame (Figure~\ref{fig:synch_attn}b), SynchFormer shifts attention to the active speaker as turns change, since its segment-based processing favors the face with the strongest audio-visual correspondence in each chunk, enabling implicit active speaker tracking without dedicated diarization or detection modules. Motivated by these properties, we adopt the frozen segment-level visual features of SynchFormer as the holistic visual condition $\mathbf{c}_v$ in CineDub.

\subsubsection{Semantic-Bundled Transcription Prompt}
\label{sec:text_condition}
However, we further observe that the attention-switching behavior of SynchFormer is not fully reliable. As illustrated in Figure~\ref{fig:synch_attn}c, attention occasionally drifts to a non-speaking face, oscillates at turn boundaries, or becomes ambiguous when speakers overlap. Additionally, when transcripts from all speakers are naively concatenated into a single text condition, the model suffers from cumulative speaker-utterance drift: slight misalignments in speaking rate or turn boundaries cause one speaker to consume the lines of another, leading to progressive collapse. Frequent turn-taking, off-screen speech, and speaker overlap further exacerbate this problem, as the holistic visual features $\mathbf{c}_v$ alone cannot anchor each utterance to its corresponding speaker.

To address this, we design a semantic-bundled transcription format as a complementary textual condition (Figure~\ref{fig:framework}, right), coupling speaker identity with utterance content at the segment level to provide additional semantic cues for speaker-utterance grounding. Given a video with $K$ utterance segments in chronological order, the transcription is structured as:
\begin{equation}
\mathbf{P} = \bigoplus_{k=1}^{K} \left[ \mathbf{d}_k \,\|\, \mathbf{u}_k \right],
\label{eq:prompt_format}
\end{equation}
where $\mathbf{d}_k$ is a natural-language description of the speaker's visual appearance or vocal characteristics, $\mathbf{u}_k$ is the corresponding transcript enclosed by special delimiter tokens \texttt{<S>} and \texttt{<E>} to distinguish spoken content from the description, $\|$ denotes within-segment concatenation, and $\bigoplus$ denotes sequential composition. Here $K$ counts utterance segments rather than speakers: consecutive segments may belong to different speakers during turn-taking, or to the same speaker when the emotion or vocal style shifts within a turn. This coupling serves as an implicit grounding mechanism: the description $\mathbf{d}_k$ provides visual or vocal cues that guide the DiT to attend to the correct face in $\mathbf{c}_v$, while the delimited transcript $\mathbf{u}_k$ is bound to that speaker's temporal trace, establishing the speaker-utterance correspondence that a flat transcript cannot provide.

In Figure~\ref{fig:framework}, the entire semantic-bundled transcription $\mathbf{P}$ is encoded by a pre-trained Gemma-T5 encoder~\cite{zhang2025t5gemma} into the unified textual condition $\mathbf{c}_t \in \mathbb{R}^{T_t \times D_t}$. A single encoder jointly represents both speaker descriptions and transcript content, requiring no separate modules. The resulting embeddings preserve sufficient lexical information for faithful speech synthesis, while retaining the semantic cues in $\mathbf{d}_k$ for speaker-face association with $\mathbf{c}_v$ and segment-level control over emotion, timbre, and vocal style. Unlike conventional dubbing pipelines, this representation bypasses explicit phoneme conversion, duration prediction, and LLM-based semantic token sequence prediction entirely, as temporal alignment is learned implicitly from the coupling of $\mathbf{c}_v$ and $\mathbf{c}_t$ during training. The format is also highly \textit{scalable}: it can be directly generated by MLLMs (e.g., Gemini 2.5 Pro~\cite{team2025gemini}), removing the need for ASR, forced alignment, and speaker diarization, and enabling annotation of large-scale in-the-wild multi-speaker datasets at minimal cost.

\vspace{5pt}
\subsection{{\fontsize{10.5}{13}\selectfont\bfseries Video to Speech and Audio Joint Generation}}
\label{sec:joint_gen}

Since $\mathbf{c}_v$ already encodes both event-level audio-visual correspondences and fine-grained lip synchronization cues, it naturally serves as a unified temporal foundation for joint speech-and-audio generation. We therefore extend CineDub to produce both within a single model. Beyond $\mathbf{c}_t$, we introduce an optional global semantic condition $\mathbf{c}_a \in \mathbb{R}^{T_a' \times D_a'}$, derived either from frame-level CLIP features at 8\,fps or from a Flan-T5 encoded natural-language audio prompt~\cite{jiang2025controlaudio} (e.g., \textit{``birds chirping with distant traffic''}). The model is then conditioned on the triplet $(\mathbf{c}_v, \mathbf{c}_t, \mathbf{c}_a)$ to produce a waveform containing both speech and ambient sound. Achieving high-fidelity joint generation, however, is non-trivial; we identify two key challenges and address each below.

\vspace{10pt}
\subsubsection{Ambient-to-Linguistic Curriculum Learning}
\label{sec:curriculum}

Under a unified DiT V2SA model with the ICHC paradigm, all generation tasks share the same parameters $\theta$ and holistic visual condition $\mathbf{c}_v$, yet differ in their target and conditioning:
\begin{equation}
\mathcal{L}_{\tau} = \mathbb{E}_{t,\epsilon}\!\left\| \epsilon_\theta(x_t^{\tau},\, \mathbf{c}^{\tau},\, t) - \epsilon \right\|^2, \quad \tau \in \{\text{a},\, \text{s},\, \text{j}\},
\label{eq:task_loss}
\end{equation}
with V2A task~$\text{a}$ setting $\mathbf{c}^{\text{a}}\!=\!(\mathbf{c}_v, \mathbf{c}_a)$ with target $x_0^{\text{a}}$, video dubbing task~$\text{s}$ setting $\mathbf{c}^{\text{s}}\!=\!(\mathbf{c}_v, \mathbf{c}_t)$ with target $x_0^{\text{s}}$, and V2SA task~$\text{j}$ setting $\mathbf{c}^{\text{j}}\!=\!(\mathbf{c}_v, \mathbf{c}_t, \mathbf{c}_a)$ with target $x_0^{\text{j}}$. The joint target $x_0^{\text{j}}$ encompasses both naturally co-occurring speech-and-sound recordings filtered from VGGSound~\cite{chen2020vggsound} and AudioSet~\cite{gemmeke2017audio}, and synthetically augmented samples where isolated speech and audio sources are mixed at controlled energy ratios to increase data diversity. 

In our early experiment (Table~\ref{tab:abl_curriculum}), we observe that naive joint training on all tasks from scratch consistently degrades performance on subtasks $\text{a}$ and $\text{s}$, particularly $\text{a}$. We attribute this to fundamental asymmetries in both the target space and the conditioning domain. On the target side, speech output is lexically anchored to transcript tokens, forming a tightly constrained mapping, whereas ambient audio occupies a far broader generative space and includes off-screen events that lack direct visual grounding. On the conditioning side, audio generation attends to diverse audio-visual events across the full visual scene, while speech concentrates on a narrow subregion (e.g., the talking face) for phoneme-level alignment. These asymmetries cause the tasks to converge at misaligned rates, leading to gradient competition, premature stopping, and catastrophic forgetting over the shared parameters $\theta$.

To resolve this, we propose an Ambient-to-Linguistic Curriculum Learning (ALC) that trains the model from the broader task to the narrower one:
\begin{equation}
\underbrace{\min_\theta \; \mathcal{L}_{\text{a}}}_{\text{Stage 1: ambient perception}} \;\longrightarrow\; \underbrace{\min_\theta \; \mathbb{E}_{\tau \sim \{\text{a},\, \text{s},\, \text{j}\}}\!\left[\mathcal{L}_{\tau}\right]}_{\text{Stage 2: linguistic specialization}}.
\label{eq:curriculum}
\end{equation}
Stage~1 trains the model exclusively on the audio generation task ($\mathcal{L}_{\text{a}}$), establishing a robust generative prior over diverse natural sounds and broad visual scene understanding. Stage~2 jointly samples all three tasks: $\mathcal{L}_{\text{a}}$ maintains the audio prior, $\mathcal{L}_{\text{s}}$ introduces speech generation, and $\mathcal{L}_{\text{j}}$ trains the model to produce coherent mixed scenes where speech and ambient sound coexist. Because the model has already learned to attend to the entire scene, narrowing its focus to the lip subregion for speech constitutes a \textit{specialization} of an existing capability rather than a conflicting objective, enabling efficient speech acquisition without disrupting the audio prior. The reverse order would be counterproductive: the model would first overfit its visual attention to the lip subregion, and subsequent audio training would both disrupt this focused attention and degrade the learned speech patterns through catastrophic forgetting (see Section~\ref{sec:ablation_curriculum} for validation).

Beyond resolving the optimization conflict, this curriculum brings two additional benefits. First, audio pre-training substantially improves paralinguistic vocalizations such as laughter, sighs, and breathing, enhancing the vocal expressiveness of dubbed characters. Second, the shared generative prior encourages acoustic consistency between synthesized speech and the ambient soundscape in energy and reverberation.


\vspace{10pt}
\subsubsection{Decoupled Textual Branch Control}
\label{sec:decoupled_attn}

With curriculum learning, a second challenge persists in joint generation: concatenating the heterogeneous conditions $\mathbf{c}_t$ and $\mathbf{c}_a$ into a single, overlong key-value sequence for a shared cross-attention module causes attention dilution. The model disproportionately attends to $\mathbf{c}_t$ at the expense of $\mathbf{c}_a$, producing faithful speech but poor adherence to the target sound effects.

We address this with a decoupled textual branch control mechanism~\cite{jiang2026freesonic} that routes the audio semantic and transcription conditions through independent branches. Within each DiT block we replace the single cross-attention layer with two parallel branches:
\begin{equation}
\mathbf{h}_{\text{cross}} = \text{CrossAttn}(\mathbf{h}, \mathbf{c}_t) + \text{CrossAttn}(\mathbf{h}, \mathbf{c}_a),
\label{eq:decoupled_attn}
\end{equation}
where $\mathbf{h}$ denotes the hidden states from the preceding self-attention layer. Because each branch maintains independent projection parameters, the model can develop specialized attention patterns for linguistic and acoustic conditions without mutual interference.

A further issue arises during single-task inference, where the inactive branch receives no meaningful input. When the absent condition is simply zeroed out, cross-task interference manifests as observable artifacts: speech-only generation produces spurious background noise from the audio pathway, while audio-only generation exhibits random off-screen voice-overs leaked from the speech pathway. To suppress this interference, we introduce learnable meta-tokens $\mathbf{m}_t$ and $\mathbf{m}_a$ as structured placeholders for the missing condition:
\begin{equation}
\mathbf{h}_{\text{cross}}^{\text{speech}} = \text{CrossAttn}(\mathbf{h}, \mathbf{c}_t) + \text{CrossAttn}(\mathbf{h}, \mathbf{m}_a),
\label{eq:expert_speech}
\end{equation}
\begin{equation}
\mathbf{h}_{\text{cross}}^{\text{audio}} = \text{CrossAttn}(\mathbf{h}, \mathbf{m}_t) + \text{CrossAttn}(\mathbf{h}, \mathbf{c}_a).
\label{eq:expert_audio}
\end{equation}
Each meta-token is a learnable embedding optimized end-to-end with the rest of the model. By occupying the inactive branch with a learned placeholder instead of a zero vector, the meta-token effectively isolates the two generation pathways, activating an expert mode that eliminates cross-task leakage and matches the performance of independently trained specialists.
\subsection{CineDub-Multi and CineDub-SA}
\label{sec:benchmark}

As noted in Section~\ref{sec:related_benchmarks}, no existing benchmark covers multi-speaker dialogue dubbing or provides joint V2SA evaluation. We construct two complementary in-the-wild benchmarks to fill these gaps.

\textbf{CineDub-Multi} is an English video dubbing benchmark designed for multi-speaker dialogue scenarios. It is derived from SpeakerVid-5M~\cite{zhang2025speakervid5m} and comprises 139 samples drawn from 139 distinct YouTube channels to maximize source diversity. Each clip is indexed by its YouTube ID and start timestamp, enabling traceability and preventing data leakage. We prioritize complex speaker interactions during curation: every clip contains at least two active speakers engaged in multi-turn dialogue, and scenes with multiple speakers visible in a single frame are included. To enforce audio-visual correspondence, clips consisting purely of narration-style audio are excluded, though brief off-screen speech caused by camera cuts is retained to reflect realistic dubbing conditions. We require clean speech without overlap or background music, while permitting natural non-linguistic vocalizations such as laughter and breathing. After filtering, we use Gemini 2.5 Pro to generate the semantic-bundled transcription (Section~\ref{sec:text_condition}) for each clip, providing speaker-utterance associations paired with per-segment speaker descriptions. All generated annotations are manually verified to ensure accuracy.

\textbf{CineDub-SA} targets V2SA evaluation, where both speech fidelity and sound-effect quality must be assessed jointly. Starting from the VGGSound test set, we apply a strict filtering pipeline that requires each clip to satisfy three criteria: a clearly visible face with corresponding speech, English spoken content (with non-linguistic vocalizations such as laughter permitted), and the co-occurrence of sound effects alongside speech. The resulting 562 ten-second clips provide verified audio-visual correspondence and are duration-matched to the segments on which standard embedding-based audio metrics are trained, enabling reliable in-domain evaluation. Each sample is accompanied by the manually verified semantic-bundled transcription and audio prompts generated by Gemini 2.5 Pro.


\begin{table*}[t]
\centering
\setlength{\aboverulesep}{1pt}
\setlength{\belowrulesep}{1pt}
\setlength{\tabcolsep}{2.0mm}
\caption{Single-speaker video dubbing results on GRID and CHEM. Methods are grouped into \emph{Hierarchical} methods that rely on face or lip crops and \emph{Holistic} methods that operate on full-frame video.
  $^*$~denotes the more challenging zero-shot voice cloning setting.
  \textcolor{gray}{Gray} entries are quoted from original papers without released checkpoints and are listed unranked for reference only.
  }
  \vspace{5pt} 
\label{tab:single_speaker}
{\footnotesize
\resizebox{\textwidth}{!}{%
\begin{tabular}{cl cccccc cccccc}
\toprule
& & \multicolumn{6}{c}{\textbf{GRID}} & \multicolumn{6}{c}{\textbf{CHEM}} \\
\cmidrule(lr){3-8} \cmidrule(lr){9-14}
& Method
  & WER\,$\downarrow$ & SIM\,$\uparrow$ & MCD\textsubscript{DTW}\,$\downarrow$ & MCD\textsubscript{SL}\,$\downarrow$ & LSE-D\,$\downarrow$ & LSE-C\,$\uparrow$
  & WER\,$\downarrow$ & SIM\,$\uparrow$ & MCD\textsubscript{DTW}\,$\downarrow$ & MCD\textsubscript{SL}\,$\downarrow$ & LSE-D\,$\downarrow$ & LSE-C\,$\uparrow$ \\
\midrule
& GT
  & 13.59 & 1.00 & 0.00 & 0.00 & 7.43 & 6.20
  & 1.44 & 1.00 & 0.00 & 0.00 & 6.67 & 7.99 \\
\midrule
\multirow{6}{*}{\rotatebox[origin=c]{90}{Hierarchical}}
& HPMDubbing~\cite{cong2023hpmdubbing}
  & 41.68 & 0.86 & 6.91 & 7.02 & 9.24 & 5.71
  & 27.39 & 0.86 & 7.03 & 8.40 & 9.63 & 4.15 \\
& StyleDubber~\cite{cong2024styledubber}
  & 17.34 & 0.91 & 6.18 & 6.57 & 8.93 & 6.23
  & 12.10 & 0.91 & 6.15 & 6.24 & 11.09 & 3.89 \\
& Speak2Dub~\cite{zhang2024speaker2dubber}
  & 14.87 & 0.92 & 6.04 & 6.59 & 9.78 & 5.34
  & 15.41 & 0.89 & 7.88 & 8.29 & 11.02 & 3.64 \\
& EmoDubber~\cite{cong2024emoDubber}
  & 16.58 & \underline{0.93} & \underline{4.43} & \underline{4.44} & \textbf{6.83} & \textbf{7.14}
  & 12.27 & 0.92 & 5.77 & 5.79 & 6.96 & \underline{7.93} \\
& \textcolor{gray}{VSSFlow-L~\cite{cheng2025vssflow}}
  & \textcolor{gray}{16.2} & \textcolor{gray}{{--}} & \textcolor{gray}{5.73} & \textcolor{gray}{5.73} & \textcolor{gray}{8.18} & \textcolor{gray}{6.81}
  & \textcolor{gray}{9.2} & \textcolor{gray}{{--}} & \textcolor{gray}{4.88} & \textcolor{gray}{4.88} & \textcolor{gray}{6.73} & \textcolor{gray}{7.89} \\
& AlignDiT$^{*}$~\cite{choi2025aligndit}
  & 23.57 & 0.91 & 5.71 & 5.72 & 7.30 & 6.50
  & 7.22 & \textbf{0.95} & 6.06 & 6.08 & 7.46 & 7.40 \\
\midrule
\multirow{4}{*}{\rotatebox[origin=c]{90}{Holistic}}
& DeepDubber~\cite{zheng2025deepdubber}
  & 54.83 & 0.74 & 13.92 & 13.94 & 10.76 & 2.82
  & 42.83 & 0.80 & 15.04 & 15.07 & 12.24 & 2.17 \\
& DeepAudio~\cite{zhang2025deepaudiov1}
  & 19.18 & 0.92 & 7.35 & 7.40 & 11.72 & 2.94
  & 25.93 & 0.92 & 8.32 & 8.36 & 13.26 & 2.29 \\
& \cellcolor{cyan!15}\textbf{CineDub$^*$} (ours)
  & \cellcolor{cyan!15}\textbf{10.36} & \cellcolor{cyan!15}\underline{0.93} & \cellcolor{cyan!15}5.08 & \cellcolor{cyan!15}5.09 & \cellcolor{cyan!15}\underline{7.27} & \cellcolor{cyan!15}\underline{6.55}
  & \cellcolor{cyan!15}\underline{6.80} & \cellcolor{cyan!15}\textbf{0.95} & \cellcolor{cyan!15}\underline{5.65} & \cellcolor{cyan!15}\underline{5.66} & \cellcolor{cyan!15}\textbf{6.71} & \cellcolor{cyan!15}\textbf{7.96} \\
& \cellcolor{cyan!15}\textbf{CineDub} (ours)
  & \cellcolor{cyan!15}\underline{13.27} & \cellcolor{cyan!15}\textbf{0.94} & \cellcolor{cyan!15}\textbf{4.35} & \cellcolor{cyan!15}\textbf{4.36} & \cellcolor{cyan!15}7.68 & \cellcolor{cyan!15}5.92
  & \cellcolor{cyan!15}\textbf{2.21} & \cellcolor{cyan!15}0.92 & \cellcolor{cyan!15}\textbf{5.03} & \cellcolor{cyan!15}\textbf{5.04} & \cellcolor{cyan!15}\underline{6.81} & \cellcolor{cyan!15}7.83 \\
\bottomrule
\end{tabular}
}
}
\end{table*}

\section{Experiment}

\subsection{Experimental Settings}

\subsubsection{Benchmarks}
We evaluate CineDub across four task settings:
(1)~\textit{Single-Speaker Video Dubbing}: following the standard protocol of~\cite{cheng2025vssflow}, we evaluate on \textit{GRID}~\cite{cooke2006grid}, an English corpus of 33 speakers with 1{,}000 utterances each (29{,}700 training / 3{,}291 test), and \textit{CHEM}~\cite{prajwal2020lip}, a single-speaker English dataset of chemistry lectures (6{,}240 training / 200 test);
(2)~\textit{Multi-Speaker Dialogue Dubbing}: evaluated on our CineDub-Multi benchmark;
(3)~\textit{Video-to-Audio Generation}: evaluated on the VGGSound test set, optionally conditioned on human-verified textual audio prompts from VGGSound-Omni~\cite{dai2026omni2sound};
(4)~\textit{Video to Speech and Audio Joint Generation}: evaluated on our CineDub-SA benchmark.
\vspace{-10pt}
\subsubsection{Evaluation Metrics}
\textbf{Audio Metrics.}
Following~\cite{dai2026omni2sound}, we assess audio generation along three dimensions:
\textit{Distribution Matching}: Fr\'{e}chet Distance (FD\textsubscript{VGG})~\cite{kilgour2019frechet} (VGGish), FD\textsubscript{PaSST} (PaSST~\cite{koutini2022passt}), and Kullback--Leibler (KL) divergence;
\textit{Audio Diversity}: Inception Score (IS) via PANNs~\cite{kong2020panns};
\textit{A-V alignment}: Desynchronization Score (Desync) via SynchFormer~\cite{iashin2024synchformer}, a standard metric widely adopted in recent V2A work~\cite{cheng2025mmaudio,dai2026omni2sound}, and ImageBind (IB) score~\cite{girdhar2023imagebind}, measuring the cosine similarity between video and audio embeddings.
\textbf{Speech Metrics.}
Following~\cite{cheng2025vssflow}, we evaluate speech generation along five dimensions:
\textit{Speech Intelligibility}: Word Error Rate (WER) via Whisper-large-v3~\cite{radford2023whisper}. In V2SA settings, we instead use Qwen3-ASR~\cite{shi2026qwen3asr} (WER\textsubscript{Qwen}) to robustly handle mixed speech-and-audio outputs. For multi-speaker dubbing, we report cpWER~\cite{yu2022comparative,kanda2021comparative}, which measures transcription accuracy and speaker attribution correctness;\footnote{cpWER concatenates per-speaker hypotheses and references chronologically, then identifies the speaker permutation that minimizes overall WER, penalizing both transcription errors and incorrect speaker assignments.}
\textit{Speaker Similarity}: SPK-SIM, the cosine similarity of WavLM-SV~\cite{chen2022wavlm} embeddings between reference and generated speech, following DeepAudio~\cite{zhang2025deepaudiov1};
\textit{Speech Naturalness}: UTokyo-SaruLab Mean Opinion Score (UTMOS)~\cite{saeki2022utmos}, a non-intrusive predictor of naturalness and clarity;
\textit{Acoustic Similarity}: Mel Cepstral Distortion with Dynamic Time Warping (MCD-DTW and MCD-DTW-SL) via ESPnet~\cite{watanabe2018espnet};
\textit{Lip Synchronization}: Lip Sync Error Distance (LSE-D) and Confidence (LSE-C) via SyncNet~\cite{chung2017syncnet} for single-speaker benchmarks. For multi-speaker dialogues, face detection on uncropped video yields prohibitively high failure rates; we therefore replace SyncNet with Desync via SynchFormer~\cite{iashin2024synchformer}, which operates on full-frame video.

\begin{table}[t]
\centering
\setlength{\aboverulesep}{1pt}
\setlength{\belowrulesep}{1pt}
\setlength{\tabcolsep}{1.8mm}
\caption{Multi-speaker dialogue video dubbing results on CineDub-Multi. cpWER extends WER by additionally penalizing speaker assignment errors, providing a more complete measure of correctness in multi-turn generation.}
\vspace{5pt}  
\label{tab:multispeaker}
{\footnotesize
\resizebox{\linewidth}{!}{%
\begin{tabular}{lcccc}
\toprule
Method & cpWER\,(WER)\,$\downarrow$ & UTMOS\,$\uparrow$ & MCD\textsubscript{DTW}\,$\downarrow$ & Desync\,$\downarrow$ \\
\midrule
GT                                           & 10.47\,(8.84)           & 2.54          & 0.00           & 0.219 \\
\midrule
AlignDiT~\cite{choi2025aligndit}            & 57.49\,(32.54)          & 2.75          & 8.98           & \underline{0.567} \\
DeepAudio~\cite{zhang2025deepaudiov1}    & 55.53\,(\textbf{12.67}) & \textbf{3.45} & \underline{8.15}  & 0.666 \\
FunCineForge\footnotemark~\cite{liu2026funcineforge} & \underline{43.47}\,(25.30) & \underline{3.45} & 10.47          & 0.882 \\
\midrule
\rowcolor{cyan!15}
\textbf{CineDub} (ours)                     & \textbf{13.93}\,(\underline{13.06}) & {2.76} & \textbf{8.06}  & \textbf{0.255} \\
\rowcolor{cyan!15}
\quad w/ flat transcript                  & 31.08\,(21.71)          & 2.83          & 8.81          & 0.396 \\
\bottomrule
\end{tabular}
}
}
\end{table}

\subsubsection{Implementation Details}
Our training follows the two-stage schedule described in Section~\ref{sec:curriculum}.
Stage~1 follows the training and data protocol of Omni2Sound~\cite{dai2026omni2sound}, constructing 470k high-quality video-audio-text triplets from VGGSound~\cite{chen2020vggsound} and AudioSet~\cite{gemmeke2017audio} for V2A pretraining.
Stage~2 introduces video dubbing and V2SA for multitask fine-tuning, drawing on 700\,hours of single-speaker and 400\,hours of multi-speaker video clips from SpeakerVid-5M~\cite{zhang2025speakervid5m}, together with over 100\,hours of V2SA clips containing co-occurring speech and sound effects from VGGSound and AudioSet.
For each speech clip, we query Gemini~2.5 Pro~\cite{team2025gemini} to generate semantic-bundled transcriptions following the format in Section~\ref{sec:text_condition}.
Following previous work, we evaluate voice cloning in single-speaker dubbing by using a reference speech from the same speaker with a different utterance.


\subsection{Main Results}

\paragraph{Single-Speaker Video Dubbing.}
Table~\ref{tab:single_speaker} compares CineDub with hierarchical and holistic baselines on CHEM and GRID, where $^*$ denotes a zero-shot setting that excludes all utterances of the test-time reference speaker from training. Despite operating on uncropped video without face or lip crops, CineDub surpasses all holistic baselines by a large margin and outperforms most hierarchical methods in speech intelligibility (WER), speaker similarity (SIM), and acoustic similarity (MCD). On GRID, EmoDubber retains an edge in lip-sync metrics (LSE-D/LSE-C) thanks to its explicit lip-crop conditioning, yet CineDub achieves comparable scores using only holistic features—suggesting that SynchFormer captures sufficient fine-grained lip cues for most practical settings. Under the zero-shot protocol, CineDub$^*$ still matches or exceeds the strongest in-domain hierarchical baselines on both datasets.


\begin{table}[t]
\centering
\setlength{\aboverulesep}{1pt}
\setlength{\belowrulesep}{1pt}
\setlength{\tabcolsep}{1.5mm}
\caption{Video-to-audio results on VGGSound test set. Top: V2A expert models; Bottom: Unified V2SA models. $\ast$~Evaluation without textual audio prompts. \textcolor{gray}{Gray} entries (unranked) are cited from original papers without released checkpoints.}
\vspace{5pt} 

\label{tab:v2a}
{\footnotesize
\resizebox{\linewidth}{!}{%
\begin{tabular}{l cccccc}
\toprule
Method & KL\,$\downarrow$ & FD\textsubscript{VGG}\,$\downarrow$ & FD\textsubscript{PaSST}\,$\downarrow$ & IS\,$\uparrow$ & Desync\,$\downarrow$ & IB\,$\uparrow$ \\
\midrule
ThinkSound~\cite{liu2025thinksound}            & 1.60          & 1.10          & 116.08         & 11.73          & 0.53          & 0.26 \\
Hunyuan-Foley~\cite{chen2024hunyuanfoley} & 1.74          & 2.36          & 100.53         & 11.58          & 0.57          & 0.32 \\
AudioX~\cite{tian2025audiox}                   & \underline{1.59} & 1.24       & 103.37         & \underline{14.94} & 1.23        & 0.26 \\
MMAudio~\cite{cheng2025mmaudio}                & 1.63          & 0.91          & 68.44          & 13.44          & \underline{0.49} & 0.29 \\
\textcolor{gray}{Omni2Sound~\cite{dai2026omni2sound}} & \textcolor{gray}{1.35} & \textcolor{gray}{0.53} & \textcolor{gray}{48.20} & \textcolor{gray}{15.79} & \textcolor{gray}{0.49} & \textcolor{gray}{0.34} \\
\midrule
\textcolor{gray}{DualDub$^*$~\cite{tian2025dualdub}}      & \textcolor{gray}{2.91} & \textcolor{gray}{2.29} & \textcolor{gray}{{--}} & \textcolor{gray}{11.50} & \textcolor{gray}{{--}} & \textcolor{gray}{0.24} \\
\textcolor{gray}{VSSFlow-L$^*$~\cite{cheng2025vssflow}}   & \textcolor{gray}{2.26} & \textcolor{gray}{1.12} & \textcolor{gray}{98.45} & \textcolor{gray}{12.83} & \textcolor{gray}{0.59} & \textcolor{gray}{0.30} \\
\rowcolor{cyan!15}
\textbf{CineDub$^*$} (ours) & 2.12 & \underline{0.65} & \underline{54.28} & \textbf{15.00} & \textbf{0.48} & \textbf{0.33} \\
\rowcolor{cyan!15}
\textbf{CineDub} (ours) & \textbf{1.41} & \textbf{0.53} & \textbf{49.91} & 14.59 & 0.50 & \underline{0.33} \\
\bottomrule
\end{tabular}
}
}
\end{table}

\begin{table*}[t]
\centering
\setlength{\aboverulesep}{1pt}
\setlength{\belowrulesep}{1pt}
\setlength{\tabcolsep}{2.5mm}
\caption{Video to speech and audio joint generation results on CineDub-SA. Cascaded baselines dub speech first, then mix audio generated by MMAudio~\cite{cheng2025mmaudio}.}
\label{tab:v2sa}
{\footnotesize
\resizebox{\textwidth}{!}{%
\begin{tabular}{l ccccc cccccc}
\toprule
& \multicolumn{5}{c}{\textbf{Speech Metrics}} & \multicolumn{6}{c}{\textbf{Audio Metrics}} \\
\cmidrule(lr){2-6} \cmidrule(lr){7-12}
Method
  & WER\textsubscript{Qwen}\,$\downarrow$ & UTMOS\,$\uparrow$ & MCD\textsubscript{DTW} \,$\downarrow$ & LSE-D\,$\downarrow$ & LSE-C\,$\uparrow$
  & KL\,$\downarrow$ & FD\textsubscript{VGG}\,$\downarrow$ & FD\textsubscript{PaSST}\,$\downarrow$ & IS\,$\uparrow$ & IB\,$\uparrow$ & Desync\,$\downarrow$ \\
\midrule
GT & 15.78 & 1.50 & 0.02 & 9.13 & 2.47 & 0.00 & 0.02 & 0.00 & 5.00 & 0.35 & 0.31 \\
\midrule
AlignDiT~\cite{choi2025aligndit}                & 38.98 & 1.72 & 11.33 & 10.30 & 2.10 & {--} & {--} & {--} & {--} & {--} & {--} \\
\quad{}+MMAudio~\cite{cheng2025mmaudio}      & 53.33 & 1.30 & 10.99 & 9.92 & 2.20 & 1.20 & 2.95 & 287.64 & 3.01 & 0.27 & 0.34 \\
DeepAudio~\cite{zhang2025deepaudiov1}        & 20.14 & \textbf{2.48} & 11.20 & 11.12 & 1.15 & {--} & {--} & {--} & {--} & {--} & {--} \\
\quad{}+MMAudio~\cite{cheng2025mmaudio}      & 51.92 & 1.36 & 11.58 & 9.99 & 2.14 & 1.17 & 2.13 & 226.55 & 3.29 & 0.30 & 0.31 \\
\midrule
\rowcolor{cyan!15}
\textbf{CineDub} (ours) & \textbf{18.65} & 1.70 & \textbf{9.51} & 9.10 & 2.81 & \textbf{0.76} & \textbf{1.24} & \textbf{171.60} & \textbf{4.00} & \textbf{0.34} & \textbf{0.19} \\
\rowcolor{cyan!15}
\quad w/ shared branch                     & 21.93 & 1.68 & 9.56 & \textbf{9.06} & \textbf{2.83} & 0.81 & 1.43 & 191.84 & 3.87 & 0.33 & 0.21 \\
\bottomrule
\end{tabular}
}
}
\end{table*}


\begin{table}[t]
\centering
\setlength{\aboverulesep}{1pt}
\setlength{\belowrulesep}{1pt}
\setlength{\tabcolsep}{2.2mm}
\caption{Ablation on curriculum learning strategies. Each variant is trained for joint V2SA generation and evaluated on GRID (speech) and VGGSound (audio).}
\label{tab:abl_curriculum}
{\footnotesize
\resizebox{\linewidth}{!}{%
\begin{tabular}{lccc|ccc}
\toprule
& \multicolumn{3}{c}{\textbf{Speech (GRID)}} & \multicolumn{3}{c}{\textbf{Audio (VGGSound)}} \\
\cmidrule(lr){2-4} \cmidrule(lr){5-7}
Strategy & WER↓ & LSE-D↓ & LSE-C↑ & FD\textsubscript{VGG}↓ & IS↑ & IB↑ \\
\midrule
Native Joint & 13.48 & 7.36 & 6.46 & 0.60 & 14.49 & 0.32 \\
L→A & 13.21 & 7.58 & 6.32 & 0.61 & 14.52 & 0.32 \\
\rowcolor{cyan!15}
A→L (ours) & \textbf{10.36} & \textbf{7.27} & \textbf{6.55} & \textbf{0.53} & \textbf{14.59} & \textbf{0.33} \\
\bottomrule
\end{tabular}
}
}
\end{table}

\paragraph{Multi-Speaker Dialogue Video Dubbing.}
Table~\ref{tab:multispeaker} evaluates multi-speaker dialogue dubbing on the proposed CineDub-Multi benchmark. All baselines show a large gap between WER and cpWER: they produce intelligible speech (low WER) but consistently misattribute it to the wrong speaker (high cpWER), e.g., a voice persists after a shot change or is assigned to the wrong speaker in multi-talker frames. We identify two root causes. First, hierarchical pipelines rely on active speaker detection, which breaks down under frequent shot changes and occlusions, yielding cpWERs of 57.49\% (AlignDiT) and 43.47\% (FunCineForge, with dedicated multi-speaker preprocessing). Second, holistic baselines lack any speaker-switching mechanism; DeepAudio reaches the lowest baseline WER (12.67\%) yet its cpWER of 55.53\% shows one identity dominates the entire dialogue. Both failures amplify the speaker-utterance ambiguity discussed in Section~\ref{sec:intro} and degrade temporal alignment (Desync 0.567--0.882). CineDub brings cpWER down to 13.93\%, a 68.0\% relative error reduction over the strongest baseline, while keeping Desync at 0.255, close to the GT level of 0.219. These gains come from the ICHC paradigm: by implicitly coupling holistic visual features with the semantic-bundled transcription, the model resolves speaker-utterance assignments and tracks speaker transitions without any dedicated preprocessing.

\vspace{-5pt}
\paragraph{Video-to-Audio Generation.}
Table~\ref{tab:v2a} benchmarks video-to-audio generation on the VGGSound test set~\cite{chen2020vggsound} against V2A specialists and V2SA models.
Despite being a unified speech-and-audio model, CineDub leads all other V2SA models across nearly all metrics in both prompt-conditioned and prompt-free settings, covering distribution fidelity (KL, FD\textsubscript{VGG}, FD\textsubscript{PaSST}), perceptual quality (IS), and audio-visual alignment (IB). It also remains competitive with state-of-the-art V2A specialists while substantially outperforming other unified V2SA models.
Combined with the speech results in Tables~\ref{tab:single_speaker}--\ref{tab:multispeaker}, these findings show that ALC combined with decoupled textual branch control effectively produces a unified model competitive with dedicated video dubbing and V2A specialists on their respective tasks, confirming that holistic visual conditioning provides sufficient cues for both speech and audio generation.

\vspace{-5pt}
\paragraph{Video to Speech and Audio Joint Generation.}
Table~\ref{tab:v2sa} evaluates joint V2SA generation on CineDub-SA against cascaded baselines that generate speech with a dubbing model and add audio via MMAudio~\cite{cheng2025mmaudio} through linear superposition. CineDub outperforms all baselines on nearly all speech and audio metrics.
The results reveal two weaknesses of the cascaded paradigm. First, standalone dubbing models achieve reasonable speech quality but poor lip sync, as in-the-wild clips with off-screen shots and non-frontal angles challenge models lacking lip conditioning. Second, the cascaded pipeline suffers from severe \textit{ghost speech}: the V2A model hallucinates speech-like sounds from visible speakers, degrading WER and UTMOS after mixing. CineDub avoids both issues via joint generation.

\subsection{Ablation Study}

\subsubsection{Effect of Semantic-Bundled Transcription}

Table~\ref{tab:multispeaker} reports an ablation in which the semantic-bundled transcription is replaced with a flat concatenation of all utterances, removing speaker descriptions and segment delimiters. This single change doubles cpWER from 13.93\% to 31.08\%, while WER degrades more moderately (13.06\%$\to$21.71\%). The disproportionate cpWER increase indicates that the model with flat transcripts still produces intelligible speech but routinely misattributes it across speakers—precisely the cumulative speaker-utterance drift described in Section~\ref{sec:text_condition}. Temporal alignment also deteriorates (Desync 0.255$\to$0.396), as the loss of per-segment anchoring weakens the coupling between visual cues and speech timing. These results confirm that semantic-bundled transcription is the critical enabler of implicit speaker grounding in multi-speaker dialogue dubbing.

\subsubsection{Effect of Curriculum Learning Strategy}
\label{sec:ablation_curriculum}

Table~\ref{tab:abl_curriculum} compares three training strategies under identical architecture, data, and total training steps: Native Joint trains on all tasks from the start without staging, L$\to$A first trains speech then adds audio, and our A$\to$L first builds an audio prior then specializes to speech. Our A$\to$L curriculum achieves the best performance on nearly all metrics. Both alternatives degrade speech: Joint training yields WER 13.48\% (vs.\ 10.36\%) due to optimization conflict between speech and audio under shared visual conditioning, and L$\to$A shows a similar degradation (WER 13.21\%). On the audio side, A$\to$L also leads across all metrics, including distribution fidelity (FD\textsubscript{VGG} 0.53 vs.\ 0.60--0.61) and audio-visual alignment (IB 0.33 vs.\ 0.32). These gains validate our hypothesis: establishing broad audio generation prior first allows speech acquisition to proceed as a specialization rather than a competing objective.

\vspace{-10pt}
\subsubsection{Effect of Decoupled Textual Branch Control}

In Table~\ref{tab:v2sa}, the ``w/ shared branch'' variant concatenates textual speech and audio condition into a single cross-attention sequence instead of routing them through independent branches. On CineDub-SA, this shared configuration degrades speech and audio quality. The degradation stems from attention dilution: the transcription dominates the shared attention at the expense of the audio prompt, leading to weaker sound-effect adherence. Decoupling the two conditions into independent branches allows each to develop specialized attention patterns, improving speech fidelity and acoustic prompt adherence. Together with ALC, this separation enables a single model to remain competitive with dedicated video-dubbing and V2A specialists (Tables~\ref{tab:single_speaker}--\ref{tab:v2a}).

\section{Conclusion}

This paper presents CineDub, a unified diffusion-based framework for multi-speaker dialogue dubbing that jointly generates speech and ambient audio from uncropped videos. At its core, the Implicitly-Coupled Holistic Conditioning (ICHC) paradigm couples holistic visual features with a semantic-bundled transcription format via cross-modal training, resolving speaker-utterance ambiguity without brittle preprocessing such as face cropping or speaker diarization. This simplicity enables scalable training on large-scale in-the-wild data and yields state-of-the-art results across single-speaker, multi-speaker, and video-to-audio benchmarks. Beyond the system, our work distills two design principles for joint generation under holistic visual conditioning: (1)~training should progress from the broader audio task to the narrower speech task to mitigate gradient competition, and (2)~textual conditions for speech and audio should be routed through separate cross-attention branches to prevent cross-prompt interference, even when the visual condition is shared. As future work, extending CineDub to multilingual dubbing is a promising direction.

\vspace{-2pt}

{
  \raggedright
  \hyphenpenalty=10000
  \exhyphenpenalty=10000
  \tolerance=9999
  \emergencystretch=3em
  \bibliographystyle{ieeenat_fullname}
  \bibliography{main}
}

%

\end{document}